\documentclass{sig-alternate}
\usepackage{dsfont}
\usepackage{url}
\usepackage{mathabx}
\usepackage{subcaption}
\usepackage{array,booktabs,calc}
\usepackage{enumitem}
\usepackage{multirow}
\usepackage{bbm}
\usepackage{balance}
\usepackage[lined,ruled]{algorithm2e}
\usepackage[titletoc]{appendix}

\newtheorem{theorem}{Theorem}[section]
\newtheorem{corollary}{Corollary}[theorem]

\DeclareMathAlphabet      {\mathbfit}{OML}{cmm}{b}{it}
\newcommand{\algname}{\textsc{sRec }}

\begin{document}

%
\conferenceinfo{WOODSTOCK}{'97 El Paso, Texas USA}

\title{Streaming Recommender Systems}
\numberofauthors{1}  
\author{  
\alignauthor Shiyu Chang$^1$\thanks{Work
    done while the author was an intern at Yahoo! Labs.}, Yang Zhang$^{1}$,  Jiliang Tang$^2$, \\
Dawei Yin$^2$, Yi Chang$^2$, Mark Hasegawa-Johnson$^1$, Thomas S. Huang$^1$\\
\affaddr{$^1$ Beckman Institute, University of Illinois at Urbana-Champaign, IL
61801, USA.}\\   \affaddr{$^2$ Yahoo! Labs, Yahoo! Inc. Sunnyvale, CA 94089,
USA.}\\
\email{\{chang87, yzhan143, jhasegaw, t-huang1\}@illinois.edu, \{jlt,daweiy,yichang\}@yahoo-inc.com} } 
\date{28 May 2015}


\maketitle
\vspace*{-0.5in}

\begin{abstract}
The increasing popularity of real-world recommender systems produces data continuously and rapidly, and it becomes more realistic to study recommender systems under streaming scenarios. Data streams present distinct properties such as temporally ordered, continuous and high-velocity, which poses tremendous challenges to traditional recommender systems. In this paper, we investigate the problem of recommendation with stream inputs.  In particular, we provide a principled framework termed \textsc{sRec}, which provides explicit continuous-time random process models of the creation of users and topics, and of the evolution of their interests.  A variational Bayesian approach called recursive meanfield approximation is proposed, which permits computationally efficient instantaneous on-line inference.  Experimental results on several real-world datasets demonstrate the advantages of our \algname over other state-of-the-arts. 



\end{abstract}

\section{Introduction}
\label{sect:intro}


Recommender systems help to overcome information overload by providing personalized suggestions from a plethora of choices based on the historical data. The pervasive use of real-world recommender systems such as eBay and Netflix generate massive data at an unprecedented rate. For example, more than 10 million transactions are made per day in eBay\footnote{\scriptsize \url{http://www.webretailer.com/articles/ebay-statistics.asp}} and Netflix gained more than three million subscribers from mid-March 2013 to April 2013\footnote{\scriptsize \url{https://en.wikipedia.org/wiki/Netflix}}. Such data is temporally ordered, continuous, high-velocity and time varying, which determines the streaming nature of data in recommender systems. Hence it is more realistic to study recommender systems using a streaming data framework. 


In recent years, there is a large literature exploiting temporal information~\cite{1099689,koren2010collaborative,conf/aaai/YinHXD11,1526930} and it is evident that explicitly modeling temporal dynamics greatly improves the recommendation performance~\cite{koren2010collaborative,DBLP:conf/sdm/XiongCHSC10}. However, the vast majority of those systems does not consider the input data as streams. Recommendation under streaming settings needs to tackle the following challenges simultaneously: 

\vspace*{0.05in}
\noindent{\bf Real-time updating}: One inherent characteristic of data streams is their high velocity; hence the recommender system needs to update and response instantaneously in order to catch users' instant intention and demands.

\vspace*{0.05in}
\noindent{\bf Unknown size}: New users or fresh posted items arrive continuously in data streams. For example, there were more than 21 million new products offered on the main Amazon USA websites from December 2013 to August 2014\footnote{\scriptsize \url{http://export-x.com/2014/08/14/many-products-amazon-sell-2/}}. Hence the number of users and the size of recommendation lists are unknown in advance. Nevertheless, many existing algorithms\cite{koren2008factorization} assume the availability of such information.

\vspace*{0.05in}
\noindent{\bf Concept shift}: Data stream evolution leads to concept shifts, \emph{e.g.}, a new product launch reduces the popularity of previous versions. Likewise, user preferences drift over time.  The recommender system should have the ability to capture such signals and timely adapt its recommendations accordingly. 

\vspace*{0.05in}
This paper defines a streaming recommender system with real-time update for a shifting concept pool of unknown size. The input streams are modeled as three types of events: user feedback activities, new users and new items. The system continuously updates its model to capture dynamics and pursue real-time recommendations.  In particular we tackle all three challenges simultaneously by a novel streaming recommender system framework (\textsc{sRec}). The major contributions are three-folds. 
\begin{itemize}[leftmargin=*]
\vspace*{-0.05in}
\item We provide a principled way to handle data as streams for effective recommendation. 
\vspace*{-0.05in}
\item We propose a novel recommender system \algname, which captures temporal dynamics under steaming settings and make real-time recommendations. 
\vspace*{-0.05in}
\item We conduct extensive experiments on various real-world datasets to understand the mechanism of the proposed framework. 
\end{itemize}




\section{Model Formulation}
\label{sect:prob_def}

We model the real-world streaming settings of recommender systems by assuming a set of continuously time-varying partially known user-item relationships as $\{R_{ij}^t\}$, where each $R_{ij}^t$ represents the rating from user $i$ to item $j$ at a given time $t$, which is modeled as a random variable. We assume that time is continuous, $t \in \mathbb{R}_+$.  In addition, $m^t \in \mathbb{Z}_+$ and $n^t \in \mathbb{Z}_+$ denote the number of users and items at time $t$. The numbers of users and items are dynamically changing over time.   

Moreover, these time-dependent ratings are partially observed. We denote $r_{ij}^t$ as the deterministic observed value of the random variable $R_{ij}^t$; $\mathcal{R}$ as a set that contains these observed rating values $r_{ij}^t$; and $\mathcal{R}^{0 : t}$ as the subset of the observed ratings no later than $t$. Here, we want to emphasize that the user-item relationship at different times are considered different: $r_{ij} ^ {t} \in \mathcal{R}$ does {\bf not} imply $r_{ij} ^ {t'} \in \mathcal{R}$, $\forall t' > t$ since we assume that the user-item relationships are inherently changing.  One advantage is that we no longer assume that an item can be rated or adopted only once by a user.  In contrast, users have the flexibility to edit or even delete ratings previously recorded.  

\subsection{Modeling User-Item Relationships}
\label{subsect:user_item_real}
Ratings, discrete and ordered, are modeled by an ordered probit model \cite{greene2008econometric}, thus rating $R_{ij}^t$ is a discretization of some hidden variable $X_{ij}^t$ that depicts the ``likeness'' of user $i$ to item $j$.  Formally, we have $R_{ij}^t = k$, if $X_{ij}^t \in (\pi_k,\pi_{k+1}]$,
where $k \in \{1,2,...,K\}$, and $K$ is the total number of discretized levels. $\{\pi_k\}_{k=1}^{K+1}$ is a set of partition thresholds, which divide the real line $\mathbb{R}$ into segments where consecutive discrete scores are assigned. The boundaries are given by $\pi_1 = -\infty$, $\pi_{K+1} = \infty$.  The intuition behind such a model is that it can easily handle both explicit and implicit feedback. For instance, Netflix uses scaled ratings ranging from 1 to 5, while Last.fm recommends music to target user based on their listening behaviors. In the first case, the rating type is explicit, and we can directly set $K=5$. On the contrary, the probit model handles implicit feedback by treating each event (\emph{e.g.} listen, download, click, \emph{etc.}) at time $t$ as a binary prediction.  Then, a single threshold is sufficient. 

The hidden user-item ``likeness'' at $t$, $X_{ij} ^t$, is modeled as
\begin{equation}
\small
\label{eq:def_X}
X_{ij} ^t = (U_i ^t)^T V_j ^t + E_{ij} ^t,
\end{equation}
where $U_i ^t$ is a $d$-dimensional hidden topic vector for user $i$ at $t$. Likewise, $V_j ^t$ is a $d$-dimensional hidden topic vector for item $j$ at $t$. $E_{ij} ^t$ is a Gaussian perturbation with mean 0 and variance $\sigma_E^2$. 

\subsection{Temporal Dynamics}
\label{subsect:temporal_dynamics}

\begin{figure}[t!]
\vspace*{-0.1in}
\begin{center}
\includegraphics[width = 0.95\linewidth]{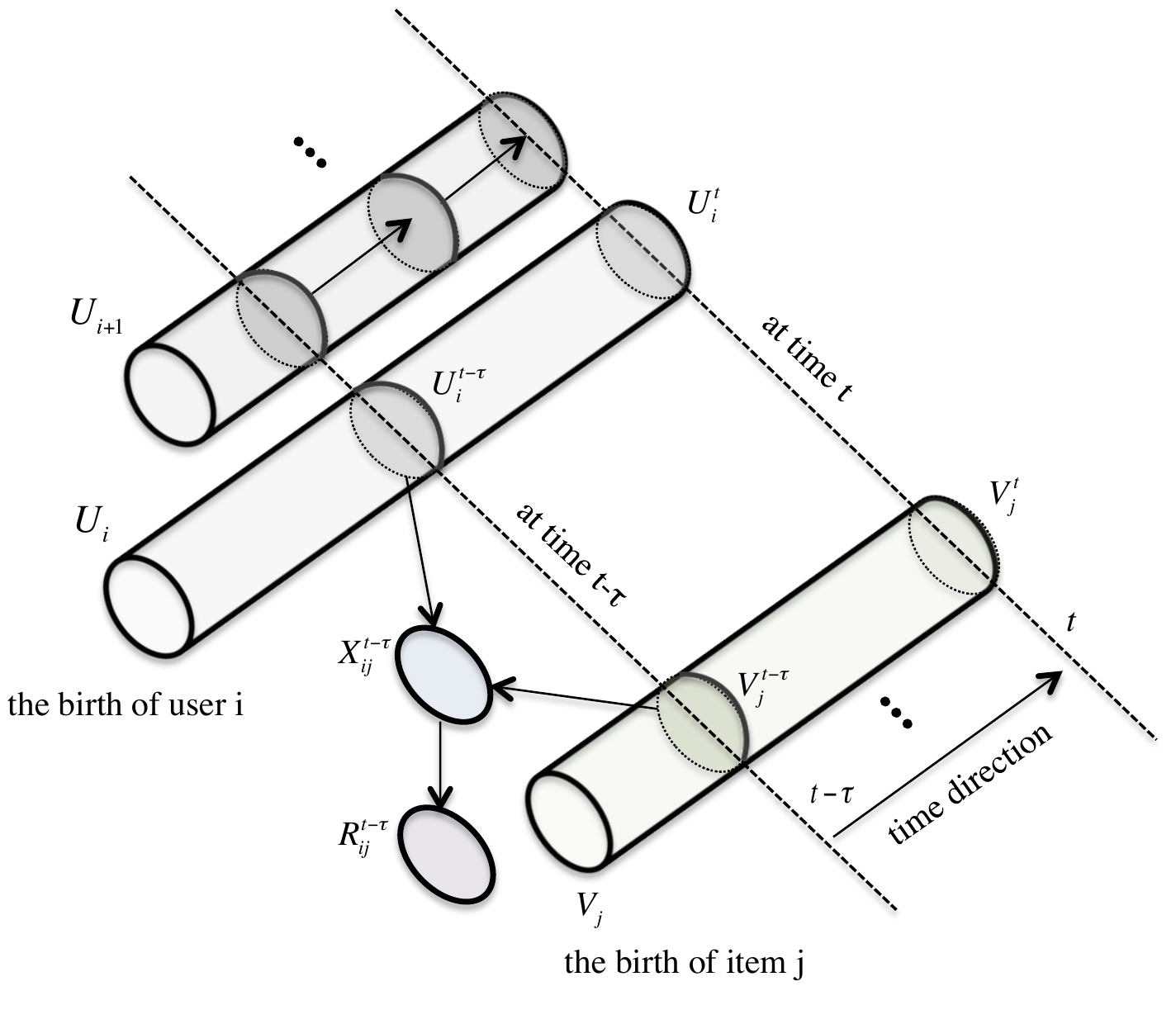}
\end{center} \vspace*{-0.3in}
\caption{Graphical illustration of the model. Each tube structure denotes the continuous Markov process of each time-varying user/item topic, with different intersections denoting the user/item topics at different times. The time axis goes up in parallel to these tubes, with the top intersections corresponding to the current time. Different users/items are born at different times, so the tubes vary in length and position of their bottom intersections. A rating that user $i$ assigns to item $j$ at time $t'$ (in the figure $t' = t - \tau$ as an example), $R_{ij} ^ {t'}$, depends on the hidden likeness $X_{ij}^{t'}$, which in turn depends on the hidden topics of user $i$ and item $j$ at time $t'$.}
\label{fig:graphical_model}
\vspace*{-0.1in}
\end{figure}

Temporal dynamics of $U_i ^t$ and $V_j^t$ incorporate both hidden topic evolution and new user/item introduction. Specifically, the hidden topic vectors of existing users follow Brownian motion
\begin{equation}
\small
\label{eq:user_dynamic}
U_{i}^t|U_{i}^{t-\tau} \sim\mathcal{N}\left(U_{i}^{t-\tau},\sigma_U^2\tau I\right), \tau > 0.
\end{equation}

For new users,  let $t_U(i)$ be the birth time of user $i$. We model the initial distribution (at the birth time), namely the distribution of $U_{i}^{t_U(i)}$, as the average of the posterior expectations of all existing user factors at $t_U(i)$ with Gaussian perturbations. We essentially assumes that the preference prior of a new user follows the general interests of others:
\begin{equation}
\small
\begin{aligned}
\label{eq:newUprob}
&U_{i}^{t_U(i)}| \left\{R_{ij} ^ t: t < t_U(i) \right\} \\
&\sim \mathcal{N}\left(\underset{k : t_U(k) < t_U(i)}{\mbox{mean}} \mathbb{E} \left(U_{k}^{t_U(i)} | \left\{R_{ij} ^ t: t < t_U(i) \right\} \right), \sigma_{U0}^2I\right).
\end{aligned}
\end{equation}
where $\underset{k : t_U(k) < t_U(i)}{\mbox{mean}}$ denotes averaging over all existing users. Furthermore, equation (\ref{eq:newUprob}) models the birth of all new users but for the very first users at $t=0$, for whom we assume a standard Gaussian prior as: 
\begin{equation}
\small
\label{eq:U0prob}
U^0_{i}\sim\mathcal{N}(0,I), \forall i, t_U(i) = 0. 
\end{equation}

Temporal dynamics as well as the initial distributions for the item factors $V_j^t$ are similar to those shown in equations (\ref{eq:user_dynamic})-(\ref{eq:U0prob}) by replacing $U_{i}^{(\cdot)}$ to $V_{j}^{(\cdot)}$, $t_U(i)$ to $t_V(j)$, and $\sigma_U^2$ and $\sigma_{U0}^2$ to $\sigma_V^2$ and $\sigma_{V0}^2$, respectively. 

\subsection{Model Summary and Notation}
\label{subsect:model_summary}
To sum up, the proposed framework consists of the following hidden variables:
\begin{equation*}
\small
\mathcal{Z} = \left\{ U_{i} ^ t, V_{j} ^ t, X_{ij} ^ t \right\} \cup \left\{R_{ij}^t : r_{ij}^t \notin \mathcal{R}\right\};
\end{equation*}
observed variables:$\left\{R_{ij}^t:r_{ij}^t \in \mathcal{R}\right\}$;
and parameters to be estimated
\begin{equation}
\label{eq:param}
\small
\Theta = \left\{\sigma_E^2, \sigma_U^2, \sigma_V^2 \right\}.
\end{equation}

For better understanding, a graphical model is shown in figure \ref{fig:graphical_model}. Moreover, before introducing the inference and training scheme for the model, we list the notations that we will use throughout the rest of the paper.

\vspace*{0.05in}
\noindent{\bf Time related notations and terminologies:}\\
\indent$\bullet~$ An event: a new user/item introduced, or a rating assigned; \\
\indent$\bullet~t$ and $t'$: : the current time and any arbitrary time, respectively;\\
\indent$\bullet~t_U(i),t_V(j)$: the birth time of user $i$ and item $j$, respectively;\\
\indent$\bullet~\tau(t)$ and $\tau'(t)$: the time between current and the most recent event time preceding and proceeding, respectively, but not including, the reference time $t$;\\
\indent$\bullet~\tau_U(i,t)$ and $\tau_V(j,t)$: the time between current and the most recent event time preceding, but not including, the reference time $t$ that involves user $i$ and item $j$, respectively;\\
\indent$\bullet~\tau'_U(i,t)$ and $\tau'_V(j,t)$: the time between current and the up-coming event time from, but not including, time $t$ for user $i$ and item $j$;\\
\indent$\bullet~\mathcal{T}$: the set of all event times;\\
\indent$\bullet~\mathcal{T}_U(i), \mathcal{T}_V(j)$: the set of all event times that are related to user $i$ and item $j$ respectively.

\vspace*{0.05in}
\noindent{\bf Random variables and distributions:}\\
\indent$\bullet~r_{ij} ^ t$: the observed value of $R_{ij} ^ t$;\\
\indent$\bullet~\mathcal{R}$: the set of values of all observed ratings $r_{ij} ^t$;\\
\indent$\bullet~\mathcal{R} ^ {0:t}$: the set of values of all observed ratings no later than $t$;\\
\indent$\bullet~\mathcal{R} ^ t$: the set of values of all observed ratings at time $t$;\\
\indent$\bullet~\mathcal{X}, \mathcal{X}^{0:t}, \mathcal{X}^t$: the set of random variables $X_{ij}^t$ whose corresponding ratings have observed values in $\mathcal{R}$, $\mathcal{R} ^ {0:t}$, $\mathcal{R} ^ t$ respectively;\\
\indent$\bullet~\bar{\mathcal{X}} ^ t$, the set of random variables $X_{ij}^t$ at time $t$ whose corresponding ratings are not observed;\\
\indent$\bullet~p_{A|B}$: the probability density function of random variable $A$ conditional on $B$, function argument omitted.

\section{Streaming Recommender System}

The proposed framework provides two components to address the following two problems: 
\begin{itemize}[leftmargin=*]
\vspace*{-0.05in}
\item {\bf Online prediction}: Based on observations UP TO the current time, how can we predict an unseen rating?
\vspace*{-0.05in}
\item {\bf Offline parameter estimation (learning)}: Given a subset of data, how can we estimate the parameters $\Theta$ defined in equation \eqref{eq:param}? 
\end{itemize}
For the first problem, we will apply posterior variational inference as shown in section \ref{subsect:online_infer}, which leads to a very efficient streaming scheme: all the predictions are based on some posterior moments, which get updated only when a relevant event happens. Furthermore, the update only depends on the incoming event and the posterior moments at the previous event, but not on other historical events.  For the second problem, we will apply variational EM algorithm, which will be derived in section \ref{subsect:offline_learning}. 


\subsection{Online Inference}
\label{subsect:online_infer}
Given past observations, for any unseen ratings {\em i.e.} $R_{ij}^t:r_{ij}^t \notin \mathcal{R} ^ {0 : t}$, the inferred rating $\hat{R}_{ij}^t$ is given by posterior expectation as:
\begin{equation}
\small
\hat{R}_{ij}^t = k, \text{ if } \mathbb{E} [X_{ij}^t | \mathcal{R}^{0 : t} ] \in (\pi_k, \pi_{k+1}].
\label{eq:infer}
\end{equation} 
The predicted rating depends on the online posterior expectation of $X_{ij}^t$, which in turn depends on posterior distributions of other hidden variables. 

\subsubsection{Recursive Meanfield Approximation}
\label{subsubsect:recur-meanfield}

According to Eq.\eqref{eq:infer}, online inference needs to evaluate the posterior distribution $p_{X^t, U^t, V^t|\mathcal{R}^{0:t}}, \forall t$, which can be decomposed into $p_{\mathcal{X}^t, U^t, V^t|\mathcal{R}^{0:t}}\cdot p_{\bar{\mathcal{X}}^t| U^t, V^t}$ by the Markov property. While the second term is readily defined by equation \eqref{eq:def_X}, it is challenging to calculate the first term due to the highly nonlinear model assumptions. Therefore, we propose a new  method, called recursive meanfield approximation, to obtain an approximate posterior distribution $q_{\mathcal{X}^{t'}, U^{t'}, V^{t'}|\mathcal{R}^{0:t'}}$, which is defined alternatively and recursively by the following three equations.

\vspace*{0.1in}
\noindent{\bf First}, $\forall t' \in \mathcal{T}$, $q_{\mathcal{X}^{t'}, U^{t'}, V^{t'}|\mathcal{R}^{0:t'}}$ is considered to be independent among $U^{t'}$, $V^{t'}$ and $X^{t'}$, and approximates the distribution induced by the Markov property
\begin{equation}
\small
\begin{aligned}
&q_{\mathcal{X}^{t'}, U^{t'}, V^{t'}|\mathcal{R}^{0:t'}}=q_{\mathcal{X}^{t'}|\mathcal{R}^{0:t'}}q_{U^{t'}|\mathcal{R}^{0:t'}}q_{V^{t'}|\mathcal{R}^{0:t'}}\\
&=\underset{q = q_{U^{t'}}q_{V^{t'}}q_{\mathcal{X}^{t'}}}{\mbox{argmin}} \mbox{ KL}\left(q_{\mathcal{X}^{t'}, U^{t'}, V^{t'}|\mathcal{R}^{0:t'-\tau(t')}}p_{\mathcal{R}^{t'} | \mathcal{X}^{t'}}\Vert q \right).
\label{eq:q-def-1}
\end{aligned}
\end{equation}

\noindent{\bf Second}, $\forall t' \notin \mathcal{T}$, since there is no event at $t'$, $\mathcal{R}^{0:t'} = \mathcal{R}^{0:t'- \tau(t')}$, thus:
\begin{equation}
\small
q_{\mathcal{X}^{t'}, U^{t'}, V^{t'}|\mathcal{R}^{0:t'}} = q_{\mathcal{X}^{t'}, U^{t'}, V^{t'}|\mathcal{R}^{0:t'- \tau(t')}}.
\label{eq:q-def-2}
\end{equation}

\noindent{\bf Finally}, both \eqref{eq:q-def-1} and \eqref{eq:q-def-2} depend on $q_{\mathcal{X}^{t'}, U^{t'}, V^{t'}|\mathcal{R}^{0:t'-\tau(t')}}$, which is defined by the Markov property:
\begin{equation}
\small
\begin{aligned}
&q_{\mathcal{X}^{t'}, U^{t'}, V^{t'}|\mathcal{R}^{0:t' - \tau(t')}}\\
=&\int q_{U^{t'-\tau(t')}|\mathcal{R}^{0:t'-\tau(t')}}q_{V^{t'-\tau(t')}|\mathcal{R}^{0:t'-\tau(t')}}p_{U^{t'}|U^{t'-\tau(t')}} \\
& \cdot p_{V^{t'}|V^{t'-\tau(t')}}p_{\mathcal{X}^{t'}|U^{t'}, V^{t'}}p_{\mathcal{R}^{t'}|\mathcal{X}^{t'}}dU^{t'-\tau(t')}dV^{t'-\tau(t')}.
\label{eq:recur-apprx}
\end{aligned}
\end{equation}

The posterior distributions at non-event times are expressed by those at the most recent event times, hence the system only needs to keep track of the posterior distributions at the most recent event time, and updates them only when an event occurs, which makes the system efficient. 

The following useful facts are stated here without proof.
\begin{theorem}
For {\bf all} $t' \in \mathcal{T}$, under recursive meanfield approximate distribution $q_{U^{t'}, V^{t'}, \mathcal{X}^{t'}|\mathcal{R}^{0:t'}}$ and \\$q_{U^{t'}, V^{t'}, \mathcal{X}^{t'}|\mathcal{R}^{0:t'-\tau(t')}}$\\
\indent $\bullet~U^{t'}$ and $V^{t'}$ follow multivariate normal distribution.\\
\indent $\bullet~$All individual user topics $U_i^{t'}$ and item topics $V_j^{t'}$ are jointly independent.
\label{thm:online-indep}
\end{theorem}
The following corollary can be derived from theorem \ref{thm:online-indep}.
\begin{corollary}
$\forall t' \in \mathcal{T}$, for user $i$ and item $j$ that are not relevant to the current ratings, i.e. $r_{ij}^{t'} \notin \mathcal{R}^{t'}$.
\begin{equation}
\small
\begin{aligned}
& \mathbb{E}_q (U_i^{t'} | \mathcal{R} ^ {0:t'}) = \mathbb{E}_q (U_i^{t'} | \mathcal{R} ^ {0:t'-\tau(t')})
= \mathbb{E}_q (U_i^{t'} | \mathcal{R} ^ {0:t'-\tau_U(i, t')}), \\
& \mbox{Cov}_q (U_i^{t'} | \mathcal{R} ^ {0:t'}) = \mbox{Cov}_q (U_i^{t'} | \mathcal{R} ^ {0:t'-\tau(t')}) + \sigma_U^2 \tau(t') I\\
=& \mbox{Cov}_q (U_i^{t'} | \mathcal{R} ^ {0:t'-\tau_U(i, t')})  + \sigma_U^2 \tau_U(t') I.
\end{aligned}
\label{eq:online-update-indep}
\end{equation}
and similarly for $V_j$.
\label{thm:online-update-indep}
\vspace*{-0.05in}
\end{corollary}

Theorem \ref{thm:online-indep} and corollary \ref{thm:online-update-indep} essentially state that 1) to keep track of the posterior distributions, we only need to keep track of their first and second moments; 2) we can update only those $U_i ^ t$ and $V_j ^ t$ that are associated with events at time $t$ and the rest are unaffected; and 3) the update does not depend on other users or items. 
The solution to the variational problem defined in \eqref{eq:q-def-1} reveals the updating equations for  $U_i^t$,$V_j^t$ and $X_{ij}^t$, $\forall t \in \mathcal{T} $. \\

\noindent {\bf Update Equations for $U_i^t$ and $V_j^t$ at Event Times:}\\
For any $U_i^t$ associated with an event at time $t$, the posterior moments are given as 
\begin{equation}
\small
\label{eq:Ui-update}
\begin{aligned}
&\mbox{Cov}_q\left(U_i^t|\mathcal{R}^{0:t}\right) \\ = &\Big[\sigma_E^{-2} \sum_{j:r_{ij}^t \in \mathcal{R}^t}\mathbb{E}_q(V_j^t(V_j^t)^T|\mathcal{R}^{0:t})+\mbox{Cov}_q(U_i^t|\mathcal{R}^{0:t-\tau( t)})^{-1}\Big]^{-1}, \\
&\mathbb{E}_q\left(U_i^t|\mathcal{R}^{0:t}\right) \\  = &\mbox{Cov}_q\left(U_i^t|\mathcal{R}^{0:t}\right)
\Big(\sigma_E^{-2} \sum_{j:r_{ij}^t \in \mathcal{R}^t} \mathbb{E}_q\left(V_j^t|\mathcal{R}^{0:t}\right) \mathbb{E}_q \left(X_{ij}^t |\mathcal{R}^{0:t}\right) \\ 
+ & \mbox{Cov}_q(U_i^t|\mathcal{R}^{0:t-\tau(t)})^{-1}\mathbb{E}_q(U_i^t|\mathcal{R}^{0:t-\tau( t)})\Big),
\end{aligned}
\end{equation}
where $\mathbb{E}_q(U_i^t|\mathcal{R}^{0:t-\tau( t)})$ and $\mbox{Cov}_q(U_i^t|\mathcal{R}^{0:t-\tau( t)})$ are given by equations \eqref{eq:online-update-indep}, \eqref{eq:newUprob}, or \eqref{eq:U0prob}, which correspond to existing users, newly born users, and very first users, respectively. The update for $V_j^t$ is identical, except that $U$ and $V$ are interchanged, and subscripts $i$ and $j$ are interchanged.

\vspace*{0.05in}
\noindent{\bf Update Equations for $X_{ij}^t$ at Event Times:}\\
On the other hand, $q_{X_{ij} ^ t|\mathcal{R}^{0:t}}$ is a truncated Gaussian distribution with the Gaussian mean
\begin{equation}
\small
\mu_{ij}^{t}=\mathbb{E}_q\left(U_{i}^{t} |\mathcal{R}^{0:t}\right)^T\mathbb{E}_q\left(V_{j}^{t} |\mathcal{R}^{0:t}\right),
\label{eq:onlineXmu}
\end{equation}
and variance $\sigma_E^2$. The truncation interval is between $(\pi_{r_{ij}^t}, \pi_{r_{ij}^t + 1}]$. Define
\begin{equation}
\vspace*{-0.05in}
\small
e_{ij}^{t}=\frac{\pi_{r_{ij}^t}-\mu_{ij}^{t}}{\sigma_E}, \text{ and } f_{ij}^{t}=\frac{\pi_{r_{ij}^t + 1}-\mu_{ij}^{t}}{\sigma_E}.
\end{equation}
The expectation of this truncated Gaussian can be expressed as
\begin{equation}
\small
\mathbb{E}_q(X_{ij}^{t} | \mathcal{R}^{0:t})=\mu_{ij}^{t}+\sigma_E\frac{\phi(e_{ij}^{t})-\phi(f_{ij}^{t})}{\Phi(f_{ij}^{t})-\Phi(e_{ij}^{t})},
\label{eq:onlineXrating}
\end{equation}
where $\phi(\cdot)$ and $\Phi(\cdot)$ are the pdf and cdf of the standard normal distribution respectively.\\

\noindent{\bf Update Equation for $X_{ij}^t$ at Non-event Times:}\\
With all the online posterior expectations derived, we are now able to obtain an approximation of $\mathbb{E}\left( X_{ij}^t  | \mathcal{R}^{0:t}\right)$, which is essential for rating prediction as in \eqref{eq:infer}:
\begin{equation}
\small
\begin{aligned}
& \mathbb{E}( X_{ij}^t  | \mathcal{R}^{0:t}) \approx \mathbb{E}_q ( X_{ij}^t  | \mathcal{R}^{0:t})
= \mathbb{E}_q ( (U_i^t )^T V_j^t  | \mathcal{R}^{0:t}) \\
= & \mathbb{E}_q(U_i^{t-\tau_U(i, t)} | \mathcal{R} ^ {0 : t-\tau_U(i, t)})^T \mathbb{E}_q (V_j^{t - \tau_V(j, t)} | \mathcal{R} ^ {0:t-\tau_V(j, t)})
\end{aligned}
\end{equation}
The last equality is given by corollary \ref{thm:online-update-indep}.

\subsubsection{Algorithm table and Order Complexity}
\begin{algorithm}[t]
\small
\SetKwInOut{Input}{input}
\SetKwInOut{Output}{output}
\Input{Current time $t$ (must be an event instance); the event type(s); currently assigned rating(s) $\mathcal{R}^t$; last updated posterior moments of $\{ U_i ^ t \}$, $\{ V_j ^ t \}$.}
\Output{Updated posterior moments of $\{U_i ^ t\}$, $\{ V_j ^ t \}$ and $\{ X_{ij} ^ t \}$ that are associated with the events at current time.}
\For{$i$ : user $i$ borns at time $t$}{
	Initialize $\mathbb{E}_q(U_i^{t} | \mathcal{R} ^ {0 : t-\tau(t)}), \mbox{Cov}_q(U_i^{t}| \mathcal{R} ^ {0 : t-\tau(t)})$ according to equations \eqref{eq:newUprob} or \eqref{eq:U0prob};
}

\For{$j$ : item $j$ borns at $t$}{
	Initialize $\mathbb{E}_q(V_j^{t}| \mathcal{R} ^ {0 : t-\tau(t)}), \mbox{Cov}_q(V_j^{t}| \mathcal{R} ^ {0 : t-\tau(t)})$ by symmetry to equations \eqref{eq:newUprob} or \eqref{eq:U0prob};
}

\While{convergence not yet reached}{
	\For{$i$ : user $i$ rates at time $t$}{
   	Update $\mathbb{E}_q(U_i^{t} | \mathcal{R} ^ {0 : t})$, $\mbox{Cov}_q(U_i^{t} | \mathcal{R} ^ {0 : t})$ according to equation \eqref{eq:Ui-update};
   }
   \For{$j$ : item $j$ is rated at time $t$}{
   	Update $\mathbb{E}_q(V_j^{t} | \mathcal{R} ^ {0 : t})$, $\mbox{Cov}_q(V_j^{t} | \mathcal{R} ^ {0 : t})$ by symmetry to equation \eqref{eq:Ui-update};
   }
   \For{$(i,j)$: user $i$ rates item $j$ at time $t$}{
   	Update $\mathbb{E}_q(X_{ij}^{t} | \mathcal{R} ^ {0 : t})$ according to equation \eqref{eq:onlineXrating}.
   }

}

\caption{Online Posterior Moments Update at Event Time}
\label{alg:posterior-update}
\end{algorithm}

To sum up, the algorithm of updating the posterior moments at event times is given in algorithm \ref{alg:posterior-update}.
The posterior moments of a user/item topic are updated only when there is an event associated with it. So the total number of updates is equal to the total number of events. In the update for each new user/item, there is a summation over all existing user/item topics of dimension $d$, which can be accelerated by setting a global sum of all the existing user/item topics. This global sum is updated only once for every event. In the update for each new rating, there is an inversion of a size-$d$ matrix, which needs no more than $O\left( d^3 \right)$ operations. Then the total computation complexity over the entire time $[0, T]$ is $O\left(I \left\lvert \mathcal{R} ^ {0 : T} \right\rvert d^3 + \left( m^t + n^t + \left\lvert \mathcal{R} ^ {0 : T} \right\rvert \right)d\right)$ where $I$ is the number iterations to reach convergence for each time.

\subsection{Offline Parameter Estimation}
\label{subsect:offline_learning}

The online streaming prediction relies on the set of parameters $\Theta$, which can be either manually predefined by domain expertise, or estimated from historical data. Define $T$ as the end time of the historical data.  This section applies expectation maximization (EM) to learn the model with incomplete observations by first defining the complete data as $(\mathcal{Z'}^{0:T},\mathcal{R}^{0:T})$ where $\mathcal{Z'}^{0:T}$ is a subset of $\mathcal{Z}$ that are associated with events, or more concretely
\begin{equation*}
\small
\begin{aligned}
\mathcal{Z'}^{0:T} = &\left\{U_i^t: t \in \mathcal{T}_U(i)\right\}\cup \left\{V_j^t: t \in \mathcal{T}_V(j)\right\} \cup \left\{X_{ij}^t: r_{ij}^t \in \mathcal{R}\right\}.
\end{aligned}
\end{equation*}

\subsubsection{M-step}
\label{subsubsect:M-step}
The auxiliary function, or Q function, is given by
\begin{equation*}
\small
Q(\Theta|\hat {\Theta}^{(k)})=\mathbb{E}_{ \mathcal{Z'}^{0:T}| \mathcal{R}^{0:T}} \left (\log p_{ \mathcal{Z'}^{0:T},  \mathcal{R}^{0:T}; \Theta}\right).
\end{equation*}
{\bf Update $\sigma_E^2$}:\\
According the first order condition, we obtain the optimal estimation formula of $\sigma_E^2$ as
\begin{equation}
\small
\label{eq:update_sigma2E}
(\hat{\sigma}_E^2)^{(k+1)}= \frac{1}{|\mathcal{R}^{0:T}|}\sum_{i,j,t': r_{ij}^{t'} \in \mathcal{R}^{0:T}} \mathbb{E}_{p} \left[\left( X_{ij}^{t'} -  U_{i}^{t'T}  V_{j}^{t'}\right)^2 |  \mathcal{R}^{0:T} ;\hat{\Theta}^{(k)}  \right],
\end{equation}
where $|\mathcal{R}^{0:T}|$ denotes the total number of ratings in the historical data from time 0 to $T$.
\vspace*{0.1in}
\\
{\bf Update $\sigma_U^2$ and $\sigma_V^2$:}\\
Similar to the $\sigma_E^2$ case, the first order condition yields
\begin{equation}
\small
\label{eq:update_sigma2U}
(\hat{\sigma}_U^2)^{(k+1)}= \frac{1}{C} \sum_{i, t' \in \mathcal{T}_U(i) \setminus t_U(i) } \frac{\mathbb{E}_{p} \left[ \| U_{i}^{t'} -  U_i^{{t'} - \tau_U(i, t')}\|^2 |  \mathcal{R}^{0:T} ;\hat{\Theta}^{(k)} \right] }{ t'-\tau_U(i,t')}.
\end{equation}
$C$ is a constant defined as the total number of user rating events. Furthermore, the estimation formula for $\sigma_V^2$ is almost identical to $\sigma_U^2$, which is omitted.
\subsubsection{E-step}
\label{subsubsect:offline-e_step}
The exact value of all offline posterior moments listed in equations (\ref{eq:update_sigma2E}) and (\ref{eq:update_sigma2U}) cannot be obtained due to the intractable inference of the posterior distributions
\begin{equation*}
\small
\begin{aligned}
p_{U^{t'},U^{t'-\tau(t')} | \mathcal{R}^{0:T}},~
p_{V^{t'},V^{t'-\tau(t')} | \mathcal{R}^{0:T}},~
p_{U^{t'},V^{t'},\mathcal{X} ^ {t'} | \mathcal{R}^{0:T}}.
\end{aligned}
\end{equation*}
However, we can utilize the result from the online inference to recursively approximate the distributions by first defining the approximated offline posterior distributions as 
\begin{equation*}
\small
\begin{aligned}
q_{U^{t'},U^{t'-\tau(t')} | \mathcal{R}^{0:T}},~
q_{V^{t'},V^{t'-\tau(t')} | \mathcal{R}^{0:T}},~
q_{U^{t'},V^{t'},\mathcal{X} ^ {t'} | \mathcal{R}^{0:T}},
\end{aligned}
\end{equation*}
which are applied to the estimation of $\sigma_U ^ 2$, $\sigma_V ^ 2$ and $\sigma_E ^ 2$ respectively.

\vspace*{0.05in}
\noindent{\bf First}, $q_{U^{t'},U^{t'-\tau(t')} | \mathcal{R}^{0:T}}$ is defined recursively by Markov property
\begin{equation*}
\small
\begin{aligned}
&q_{U^{t'},U^{t'-\tau(t')} | \mathcal{R}^{0:T}} = q_{U^{t'}| \mathcal{R}^{0:T}} q_{U^{t'-\tau(t')} |U^{t'}, \mathcal{R}^{0:t'-\tau(t')}} \\
= & q_{U^{t'}| \mathcal{R}^{0:T}} \frac{q_{U^{t'-\tau(t')} |\mathcal{R}^{0:t'-\tau(t')}} p_{U^{t'}|U^{t'-\tau(t')}}}{\int q_{U^{t'-\tau(t')} |\mathcal{R}^{0:t'-\tau(t')}} p_{U^{t'}|U^{t'-\tau(t')}} dU^{t'}},
\end{aligned}
\end{equation*}
where the first term is given backward-recursively by marginalizing $q_{U^{t' + \tau'(t')},U^{t'} | \mathcal{R}^{0:T}}$; and $q_{U^{t'-\tau(t')} |\mathcal{R}^{0:t'-\tau(t')}}$ in the fraction is given by online inference.  $q_{V^{t'},V^{t'-\tau(t')} | \mathcal{R}^{0:T}}$ is defined similarly and is omitted.

\vspace*{0.05in}
\noindent{\bf Second}, $q_{U^{t'},V^{t'},\mathcal{X} ^ {t'} | \mathcal{R}^{0:T}}$, similar to \eqref{eq:q-def-1}, assumes $U^{t'},V^{t'},\mathcal{X} ^ {t'}$ are independent
\begin{equation*}
\small
\begin{aligned}
&q_{\mathcal{X}^{t'}, U^{t'}, V^{t'}|\mathcal{R}^{0:T}}=q_{\mathcal{X}^{t'}|\mathcal{R}^{0:T}}q_{U^{t'}|\mathcal{R}^{0:T}}q_{V^{t'}|\mathcal{R}^{0:T}},
\label{eq:q-offline-def-4}
\end{aligned}
\end{equation*}
where $q_{U^{t'}|\mathcal{R}^{0:T}}$ and $q_{V^{t'}|\mathcal{R}^{0:T}}$ are already defined, and $q_{\mathcal{X}^{t'}|\mathcal{R}^{0:T}}$ is set to minimize the KL divergence to the true distribution.
\begin{equation*}
\small
\begin{aligned}
&q_{\mathcal{X}^{t'}|\mathcal{R}^{0:T}}
=\underset{q}{\mbox{argmin}} \mbox{ KL }(p_{U^{t'},V^{t'},\mathcal{X} ^ {t'} | \mathcal{R}^{0:T}} \Vert q \cdot q_{U^{t'}|\mathcal{R}^{0:T}}q_{V^{t'}|\mathcal{R}^{0:T}}).
\label{eq:q-offline-def-5}
\end{aligned}
\end{equation*}
Similar to the online case, these approximated distributions have good properties, which will be stated without proof.
\begin{corollary}
\label{cor:offline}
Under the approximated offline distribution
$q_{U^{t'}, U^{t'-\tau(t')} | \mathcal{R}^{0:T}}$ and $q_{V^{t'}V^{t'-\tau(t')} | \mathcal{R}^{0:T}}$, all individual user topics $U_i^{t'}$ and item topics $V_j^{t'}$ are independent.
\end{corollary}
Using the corollary \ref{cor:offline}, all the offline posterior expectations listed in equations (\ref{eq:update_sigma2E}) and (\ref{eq:update_sigma2U}) can be computed. 





\section{Experiments}
\label{sect:exp}


To assess the effectiveness of the proposed framework, we collect three real-world datasets and their detailed descriptions are as follows:



\vspace*{0.05in}
\noindent \textit{\textbf {MovieLens Latest Small}}: It consists of 100,000 ratings to 8,570 movies by 706 users.  In addition, all ratings are associated with timestamps ranging from the year 1996 to 2015. The ratings are given on a half star scale from 0.5 to 5. In abbreviate, we use \emph{ML-latest} for the following subsections. 

\vspace*{0.05in}
\noindent \textit{\textbf {MovieLens-10M}}: The dataset contains 10 million movie ratings from 1995 to 2009.  The rating scale and temporal information are similar to those of the \emph{ML-latest} dataset.  In total, there are 10,000 movies and 71,000 users.  We use \emph{ML-10M} to represent the dataset.

\vspace*{0.05in}
\noindent \textit{\textbf {Netflix}}: The \emph{Netflix} dataset contains 100 million movie ratings from 1999 to 2006 that are distributed across 17,7700 movies and 480,189 users.  The rating scale is from $1$ to $5$.  Furthermore, the temporal granularity is a day.





\subsection{Baseline Methods}
\label{subsect:baseline}
We compare our proposed framework \algname with several representative recommender systems including:



\noindent {\bf PMF~\cite{mnih2007probabilistic}}: Probabilistic Matrix Factorization is a classical recommendation algorithm that is widely used.  

\noindent {\bf MDCR~\cite{bhargava2015and}}: Multi-Dimensional Collaborative Recommendation is a tensor based matrix factorization algorithm, which can potentially incorporate both the temporal and spacial information.  

\noindent {\bf Time-SVD${++}$~\cite{koren2010collaborative}}: Time-variant version of the Netflix winning algorithm SVD${++}$~\cite{koren2008factorization}.  In addition, the model can be evolved efficiently via a online updating scheme.  

\noindent {\bf GPFM~\cite{nguyen2014gaussian}}: Gaussian Process Factorization Machines is a variant of the factorization machines \cite{rendle2010factorization}, which is non-linear, probabilistic and time-aware.  

In summary, PMF is a static model, and all the other three baselines leverage the temporal factor in different ways.  We utilize the existing C$++$ implementations from GraphChi \cite{kyrola2012graphchi} for PMF and Time-SVD${++}$, while the code for GPFM is also available online\footnote{\scriptsize \url{https://github.com/trungngv/gpfm} }. 

\subsection{Experimental Settings}
\label{subsect:settings}

We divide all data into halves.  We call the first half the ``base training set'', which is used for parameter estimation; and the remaining half is the ``candidate set''.  The base training set can be considered as the historical data pool while the candidate set mimics the steaming inputs.  Both the proposed method and other baseline algorithms utilize the based training set to determine the best hyper parameters.  For instance, the latent dimensionality for each algorithms are determined independently using this base training set.  For online prediction, the actual testing set is generated from the candidate set by randomly selecting a reference time first. Such a reference time can be considered as the ``current-time''. Then, our task is to predict user-item ratings for the following week starting from the reference time.  All the ratings prior to the reference time are used for the online inference.  The sequential order of data streams ensures a prospective inference procedure: no future ratings can be used to predict the past behaviors.  It is worth mentioning that, except our \algname model, other baselines cannot explicitly handle new user/item introduction during the testing phrase.  Therefore, for a fair comparison, all the testing ratings are from the existing users and items which have appeared in the training set.  

We evaluate the performance via the root mean square error (RMSE), which is a widely used metric in recommendation.  Furthermore, in order to ensure reliability, all experimental results were averaged over 10 different runs. In addition, to keep the temporal variability off the testing set, there is no temporal overlapping between any testing sets.  


\subsection{Experimental Results}
\label{subsect:results}
The best performance of each method on all three datasets is reported in table \ref{tab:result}. We observe that the proposed algorithm achieves the best performance across all three datasets. It is evident that explicitly modeling user/item dynamics significantly boosts the recommendation performance under the streaming setting. 
The second best algorithm is Time-SVD$++$, which is better than the other three baselines consistently. This is because, Time-SVD$++$ models temporal information in a more suitable way under the streaming settings.  However, the inflexibility to the temporal granularity makes it insufficient to capture any volatile changes.  On the other hand, the tensor based method MDCR yields the worst performance.  One potential reason is that the algorithm considers temporal information in a retrospective way, which is obviously inadequate to capture the prospective process of data generation.   Furthermore, the PMF method does not utilize any time information, however, its performance is still better than MDCR.  This result implies that inappropriate temporal modeling even hurts the performance.  Last, GPFM suffers from the problem of scalability, thus it provides no result for the Netflix dataset.

\begin{table}[t]
\centering
\caption{Performance comparison in terms of RMSE. The best performance is highlighted in bold.}
\label{tab:result}
\vspace*{-0.1in}
\begin{tabular}{c|c|c|c}
\hline
             & ML-Latest & ML-10M & Netflix \\ \hline
PMF          & 0.7372           & 0.8814        & 0.8610  \\ 
MDCR         & 0.7604           & 0.9349        & 0.9326  \\ 
GPFM         & 0.7710           & 0.9114        & NA      \\ 
Time-SVD$++$ & 0.7141           & 0.8579        & 0.8446  \\ 
\algname     & {\bf 0.6780}           & {\bf 0.7957}        & {\bf 0.8093}  \\ \hline
\end{tabular}
\vspace*{-0.15in}
\end{table}

\subsection{Parameter Understanding}
\label{subsect:parameter}
In this subsection, we investigate the physical meanings behind our learned parameters. The major parameters for the proposed method are $\sigma_E^2$, $\sigma_U^2$ and $\sigma_V^2$.  According to our experimental settings, these parameters are learned from the so called historical data.  Note that, in this part, we will only show the results on the \emph{ML-Latest} dataset. The results for other datasets are similar.

\begin{figure}[t]
\begin{center}
\includegraphics[width = 1\linewidth]{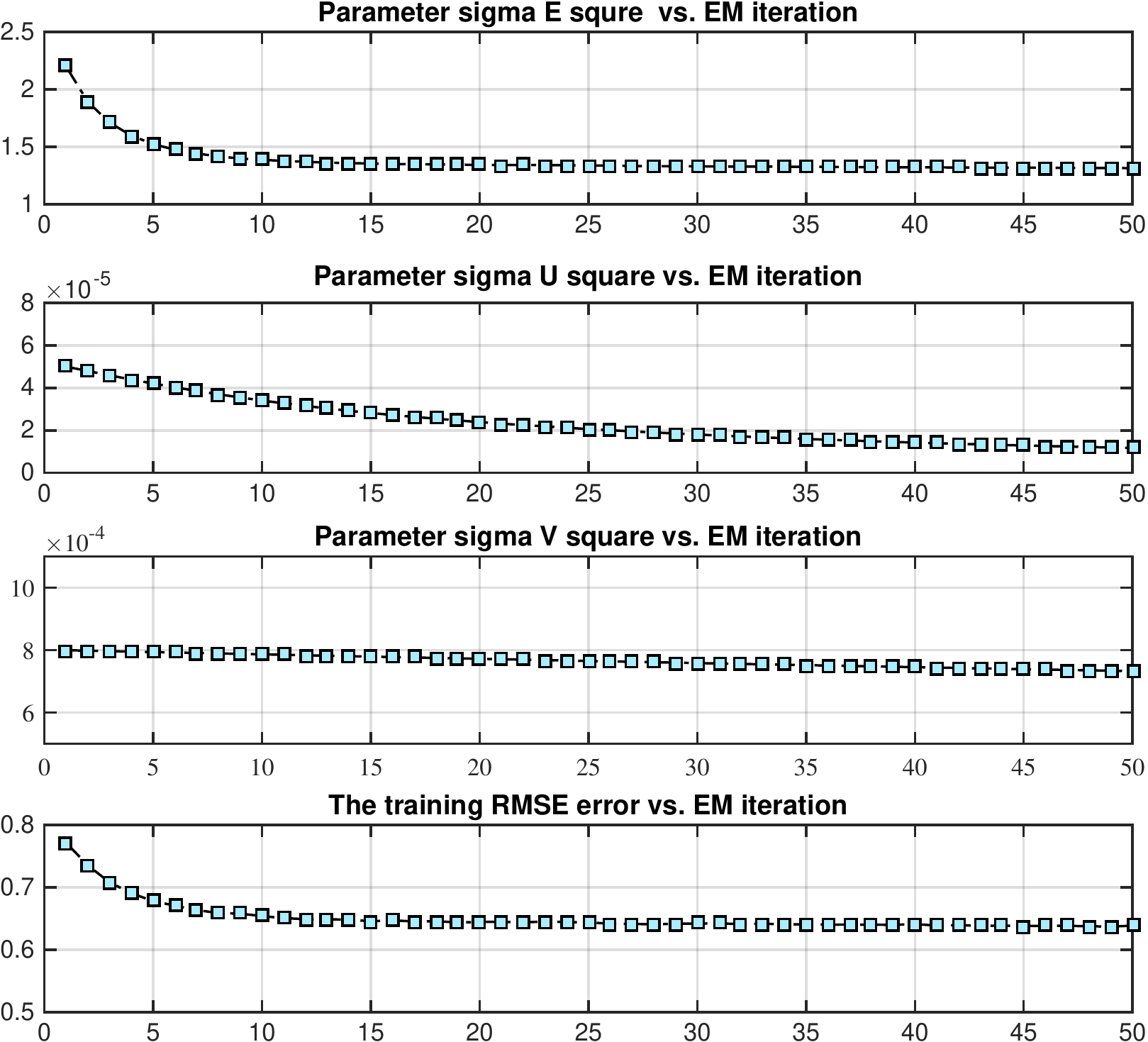}
\end{center} \vspace*{-0.25in}
\caption{The convergence analysis for learning parameters on the \emph{ML-Latest} dataset. }
\label{fig:EM_convergence}
\vspace*{-0.15in}
\end{figure}

We first perform empirical convergence analysis for the proposed \algname method.  Figure \ref{fig:EM_convergence} demonstrates the convergence paths of all three parameters as well as the training error against EM iterations.  
As we can see, all the parameters converge, and the RMSE on the training set is saturated around the 10th iteration. The converged values of $\sigma_E^2$, $\sigma_U^2$ and $\sigma_V^2$ are $1.32$, $1.17\times 10^{-5}$ and $7.33\times 10^{-4}$ respectively while the training RMSE is around $0.64$.  Next, we analyze the interpretations behind the values of these parameters.  


The first parameter can be understood as the impact of other unknown factors on ratings. The larger the value of $\sigma_E^2$, the more susceptible the rating is to variations in unknown factors, and hence the less predictable by \textsc{sRec}.

\begin{figure}[t]
\begin{center}
\includegraphics[width = 0.7\linewidth]{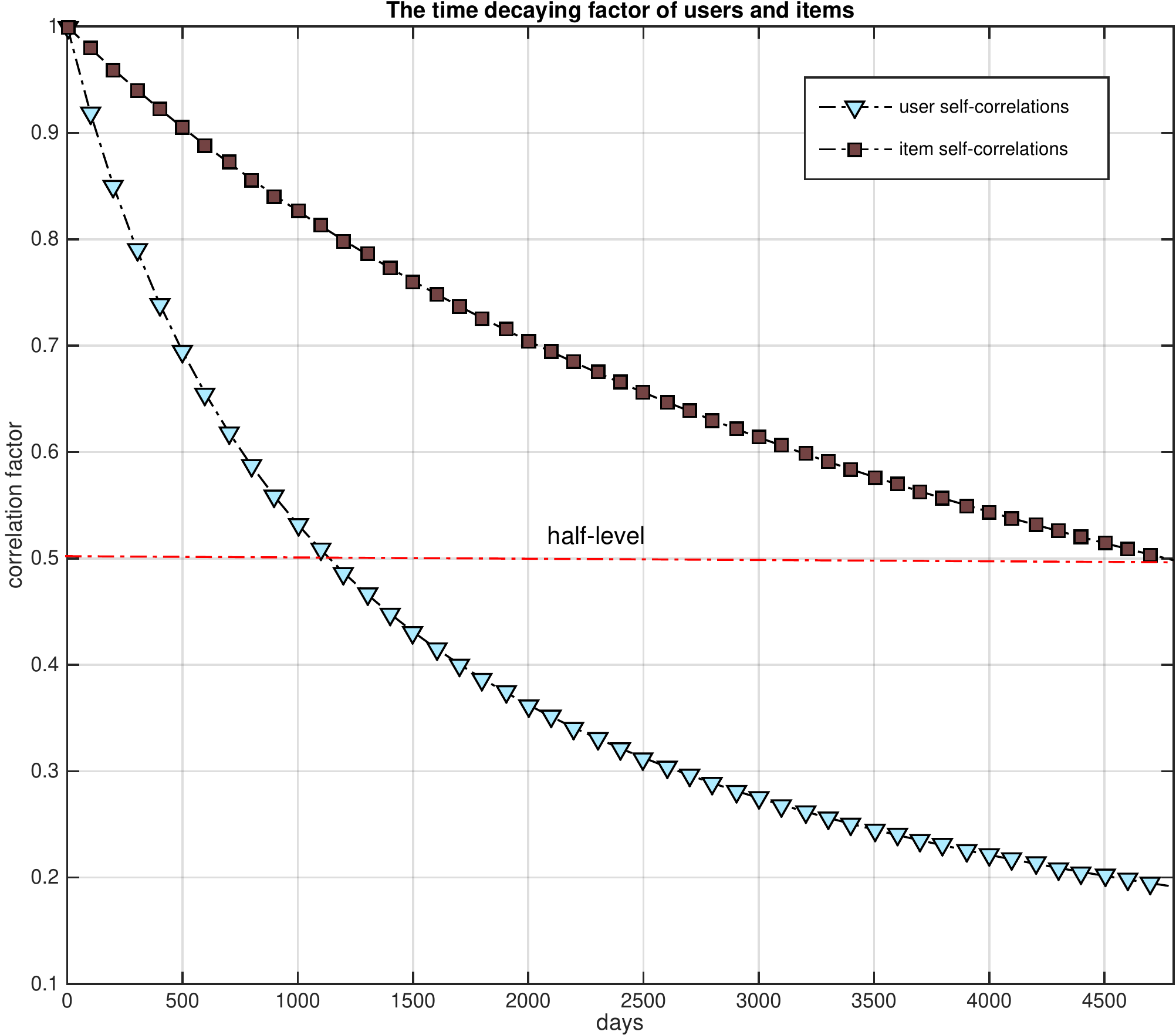}
\end{center} \vspace*{-0.25in}
\caption{Correlation Decay with time. The half-time for user and item is 3.12 years and 13.06 years respectively.}
\label{fig:half-time}
\vspace*{-0.05in}
\end{figure}

More interestingly, $\sigma_U^2$ and $\sigma_V^2$ are intuitively considered as the evolution speed of user/item taste based on our model assumptions in a global sense (across the whole population of the users / items). We relate the learned values of these two parameters to our intuitions in a quantitative way via the following posterior correlation function:
\begin{equation}
\small
\label{eq:half-time}
\begin{aligned}
\mbox{Corr}(U_i^{t'},U_i^{t'+\delta_t} | \mathcal{R}^{1:t'})^2 
= \frac{\mbox{Tr}(\mbox{Cov}(U_i^{t'} | \mathcal{R}^{1:t'} )}{\mbox{Tr}(\mbox{Cov}(U_i^{t'} | \mathcal{R}^{1:t'} ) +  d \delta_t\sigma_U^2}.
\end{aligned}
\end{equation}
This quantity is a function of $\sigma_U^2$, which measures the user auto-correlation from a reference time $t'$ to some future time $t'+\delta_t$. The range of such a measurement is from 0 to 1. The item correlation function is identical to equation (\ref{eq:half-time}) by replacing the corresponding $U$ with $V$. We vary $\delta_t$ and plot the mean of these correlations for all users and items in figure \ref{fig:half-time}. The reference time $t'$ is equal to the last time instance in the training data, and $\sigma_U^2$ and $\sigma_V^2$ are obtained from the offline learning.  

Furthermore, one critical value for the correlation measure is 0.5, which is usually referred as the ``half-time''.  In figure \ref{fig:half-time}, the half-time for both users and items are the intersections of the red dashed line to their own curves.  The``half-time'' for users is around 3.12 years; while the one for items is 13.06 years, which can be understood as: a rating provided by a user can only be trusted to reflect half his/her taste for 3.12 years. Although we indeed observe the change of item topics over time, the change is significantly slower than the user ones. This finding can be used to explain why many existing recommender systems only consider the dynamics of user topics while assuming fixed item topics.

\subsection{Temporal Drifting}
\begin{figure}[t]
\begin{subfigure}{.24\textwidth}
  \centering
  \includegraphics[width= 1 \linewidth]{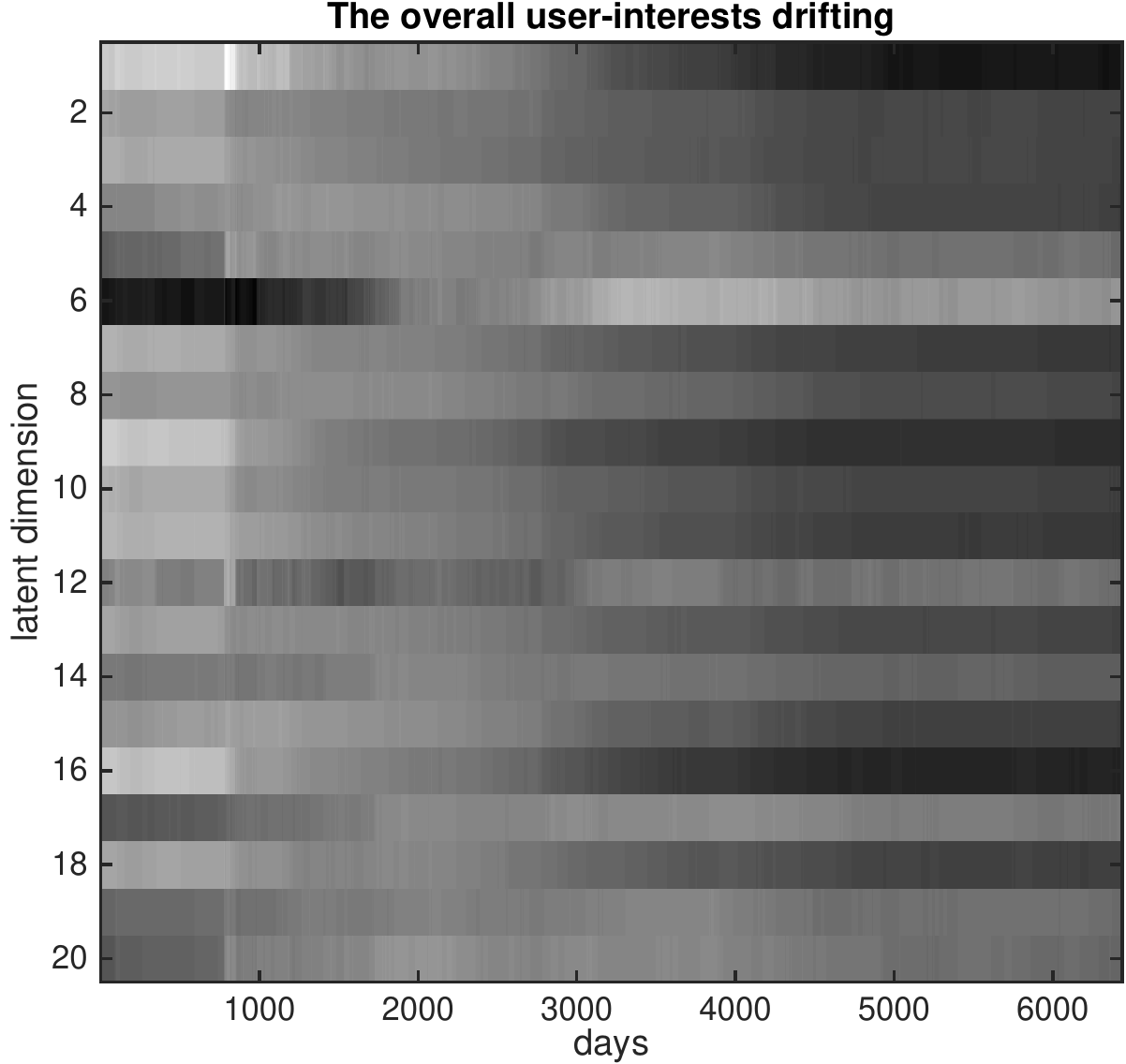}
  \caption{User}
  \label{fig:global_topic_changes_U}
\end{subfigure}%
\begin{subfigure}{.24\textwidth}
  \centering
  \includegraphics[width= 1 \linewidth]{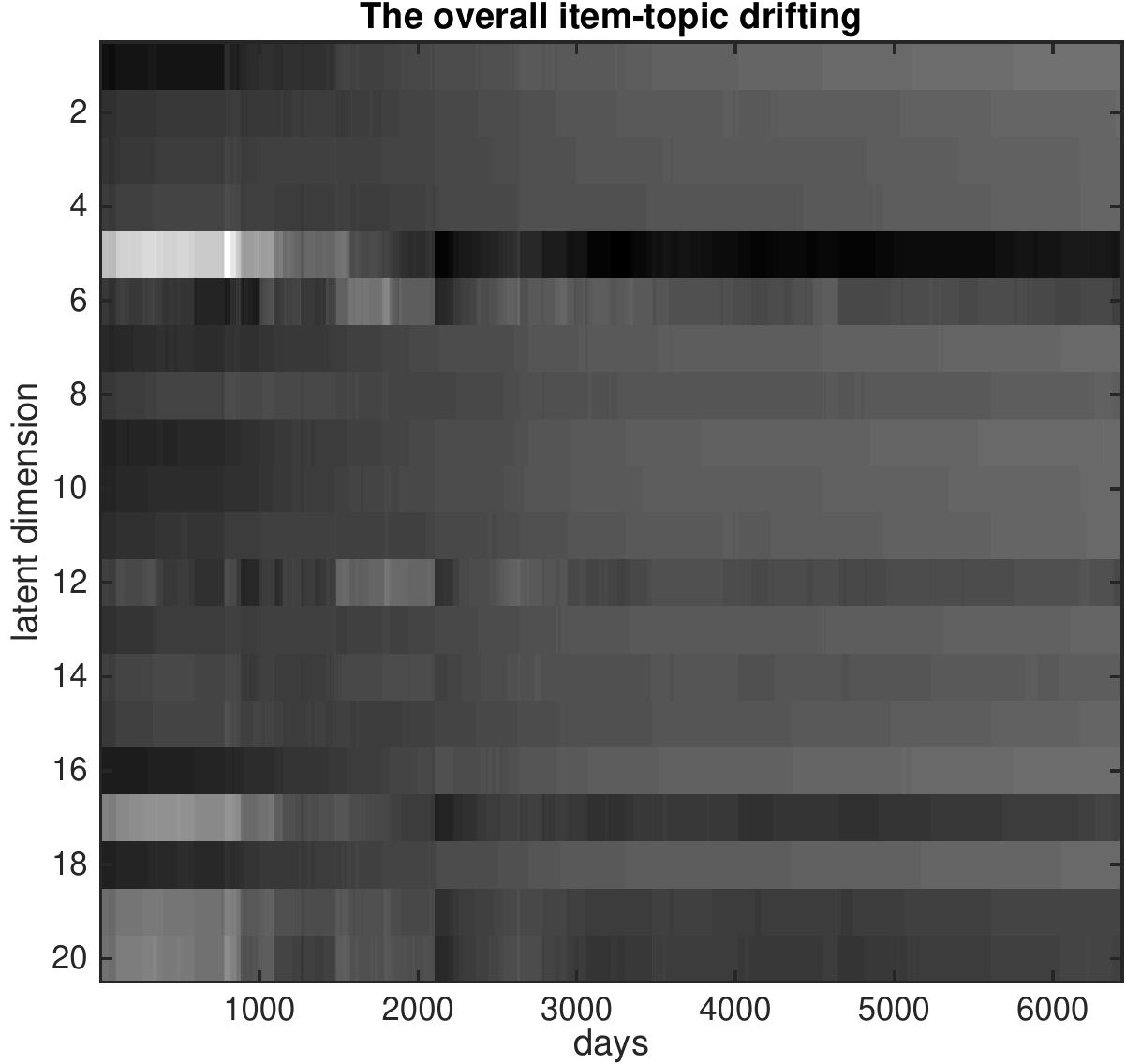}
  \caption{Item}
  \label{fig:global_topic_changes_V}
\end{subfigure}
\vspace*{-0.1in}
\caption{The averaged latent topics for both users (\ref{fig:global_topic_changes_U}) and items (\ref{fig:global_topic_changes_V}) at different time.}
\label{fig:global_topic_changes}
\vspace*{-0.1in}
\end{figure}

We demonstrate the proposed \algname can capture the temporal drifting phenomenon from another perspective. We recorded all the posterior expectations of user/item topics 
as a function of time. 
Figure \ref{fig:global_topic_changes} shows the evolution of latent factors with the \emph{ML-Latest} dataset, which x-axis denotes time and y-axis is the latent dimension.  
The intensity of the user latent representation reflects the user likeness to a specific hidden topic. The darker the color in figure \ref{fig:global_topic_changes_U}, the less emphases users lay on the certain topic.  On the other hand, the item latent vectors indicate the assignment of items to these latent topics.  The observation is consistent with our experimental results in section \ref{subsect:parameter}: the user tastes change more remarkably compared to item topics. 


Furthermore, several topics evolve much more drastically compared to others. Notable ones include the 1st, 6th, 12th and 16th user topics as well as the 5th item topic.  To understand the reason, we provide an example using the 6th user topic, which has a volatile shift in interests around the 1,500 day (year 2001).  
Table \ref{tab:top_6_topic} shows the top movies that have strongest responses at the 6th latent topics. Most of the movies are popular commercial type with the majority of their ratings provided after day 1500.  Moreover, the average rating for the most of these movies are substantially higher than that of the whole dataset, which is 3.49.  Another indicatory event is the movie ``The Lord of Rings'' that was also released in that year. This might potentially correlate to user taste changes and affect the latent representation. 

\begin{table*}[th]
\small
\centering
\caption{The statistics of the top movies that contain strong responses at their 6th latent dimension.}
\label{tab:top_6_topic}
\vspace*{-0.1in}
\begin{tabular}{|c|c|c|c|c|c|c|}
\hline
Topic 6                     & Rank 1                        & Rank 2                 & Rank 3                 & Rank 4                               & Rank 5                        & Rank 6                       \\ \hline
\multirow{2}{*}{Movie name} & \multirow{2}{*}{Start War V} & There's Something      & \multirow{2}{*}{Alien} & \multirow{2}{*}{The Lord of Rings I} & \multirow{2}{*}{Star Wars IV} & \multirow{2}{*}{Star War VI} \\
                            &                               & About Mary             &                        &                                      &                               &                              \\ \hline
\multirow{2}{*}{Genre}      & Action, Sci-Fi                & Comedy,                & Horror,                & Adventure,                           & Action, Sci-Fi                & Action, Sci-Fi               \\
                            & Adventure                     & Romance                & Sci-Fi                 & Fantasy                              & Adventure                     & Adventure                    \\ \hline
Release time                & 1980                          & 1998                   & 1979                   & 2001                                 & 1977                          & 1983                         \\ \hline
Percentage of ratings       & \multirow{2}{*}{75.43\%}        & \multirow{2}{*}{88.39\%} & \multirow{2}{*}{78.87\%} & \multirow{2}{*}{100\%}                   & \multirow{2}{*}{69.15\%}        & \multirow{2}{*}{70.46\%}       \\
after day 1500              &                               &                        &                        &                                      &                               &                              \\ \hline
Average ratings             & \multirow{2}{*}{4.17}         & \multirow{2}{*}{3.64}  & \multirow{2}{*}{4.02}  & \multirow{2}{*}{4.04}                & \multirow{2}{*}{4.14}         & \multirow{2}{*}{3.99}        \\
after day 1500              &                               &                        &                        &                                      &                               &                              \\ \hline
\end{tabular}
\vspace*{-0.05in}
\end{table*}

\begin{figure}[t]
\begin{center}
\includegraphics[width = 0.7\linewidth]{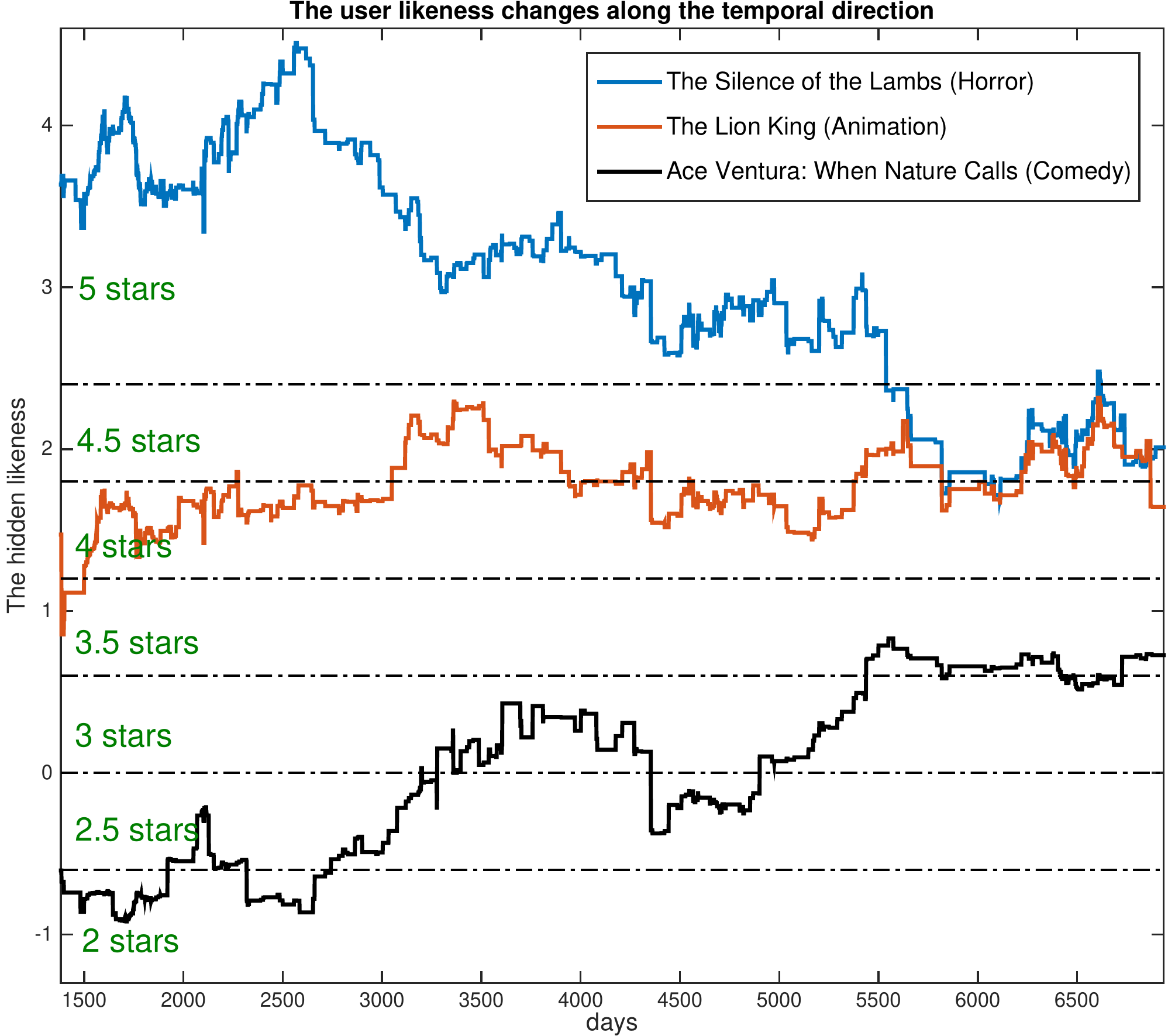}
\end{center} \vspace*{-0.25in}
\caption{An example of user interest evolving over time through predicted ratings for some representative movies.}
\label{fig:predicted_rating}
\vspace*{-0.2in}
\end{figure}

We next present a qualitative result to demonstrate the advantage of our \algname model in tracking user instant preference by plotting the predicted ``likeness'' (posterior expectation of $X_{ij}$) for one of the most active users from the data to a set of movies as a function of time. In figure \ref{fig:predicted_rating}, the preference of an individual user continuously evolves on those three selected movies.  An initial strong interest in ``The Silence of the Lamb'' decreases slightly with time.  In contrast, the comedy movie ``Ace Ventura'' becomes more and more preferable.  The last movie ``The Lion King'' is considered as a favorable movie by the user across the entire time.

\section{Related Work}
\label{sect:related_work}
\subsection{Online Recommender System}
Most previous works focus on traditional batch regression problem \cite{nguyen2014gaussian}. However, the recommender system by nature is an incremental process. That is, users' feedback are performed in a sequence,  and their interests are also dynamically evolving. To capture the system evolution, once a user feedback is performed, the recommender system should be able to get updated. Authors in \cite{Agarwal:2010:FOL:1835804.1835894} proposed a fast online bilinear factor model to learn item-specific factors through online regression, where each item can perform independent updates.  Hence, this procedure is fast, scalable and easily parallelizable. \cite{Rendle:2008:ORK:1454008.1454047} introduced regularized kernel matrix factorization method where kernels provide a flexible way to model nonlinear interactions and an online-update scheme. Das \emph{et al.}~\cite{Das:2007:GNP:1242572.1242610} addressed the large data and dynamic content problem in recommender systems, and proposed online models to generate personalized recommendations for Google News users. Diaz \emph{et al.}~\cite{Diaz-Aviles:2012:RTR:2365952.2365968} presented Stream Ranking Matrix Factorization, which uses a pairwise approach to matrix factorization in order to optimize the personalized ranking of topics and follows a selective sampling strategy to perform incremental model updates based on active learning principles. \cite{Chen:2013:TTR:2536274.2536289} extended the online ranking technique and proposed a temporal recommender system: when posting tweets, users can get recommendations of topics (hashtags) according to their real-time interests. Furthermore, users can also generate fast feedback according to the obtained recommendations. However, the above methods focus on online update and learning, but neglect the temporal ordering information of the input data. 

Recently, beyond the online learning methodology, the concept of real-time recommender system is introduced, which emphasizes on the scalability and real-time pruning in recommender systems.
Authors in \cite{Huang:2015:TRS:2723372.2742785} present a practical scalable item-based collaborative filtering (CF) algorithm, with the characteristics such as robustness to the implicit feedback problem, incremental update and real-time pruning. Zhao \emph{et. al.} \cite{zhao2013interactive} utilize a selective sampling scheme under the PMF framework \cite{mnih2007probabilistic} to achieve interactive updates.  StreamRec \cite{export:145190} is invented based on a scalable CF recommendation model, which emphasized on the databases aspect of a stream processing system.  A neighborhood-based method for streaming recommendations is discussed in \cite{2016CharuCIKM}.  The proposed framework is substantially different from the above real-time recommender systems: aforementioned systems are memory based methods and designed only for certain applications, while our method provides a principled way to handle data as streams. 

\subsection{Time-aware Recommendations}
An important factor of time-aware recommendations is how to explicitly model users' interest change over time. In collaborative filtering, modeling temporal information has already shown its success (\emph{e.g.} Ding and Li \shortcite{1099689}, Koren \shortcite{koren2010collaborative}, Yin \emph{et al.} \cite{conf/aaai/YinHXD11} and Xiang \emph{et al.} \shortcite{1835896}). Zhang \emph{et al.}\cite{1526930} investigated the recurrence dynamics of social tagging.  
Xiong \emph{et al.}~\cite{DBLP:conf/sdm/XiongCHSC10} used tensor factorization approach to model temporal factor, where the latent factors are under Markovian assumption. Bhargava \emph{et al.}~\cite{bhargava2015and} further extended the tensor factorization to handle not only temporal but also geographical information. Liang \emph{et al.}~\cite{Liang:2012:TTR:2396761.2398492} proposed to use the implicit information network, and the temporal information of micro-blogs to find semantically and temporally relevant topics in order to profile user interest drifts.  Liu \emph{et al.}~\cite{Liu:2013:STC:2414425.2414440} extended collaborative filtering to a sequential matrix factorization model to enable time-aware parameterization with temporal smoothness regularization. Both \cite{gultekin2014collaborative} and \cite{lu2009spatio} model the temporal dynamics using Kalman Filtering, which is somewhat similar to our online prediction framework.  However, we differ from their method by providing more principled inference procedures.  In addition, we propose a data-driven technique to automatic estimate all the parameters that have physical meanings (\emph{i.e.} the half-time).  Furthermore, as the temporal factors catching more attention, Burke \emph{et al.}~\cite{Burke:2010:EDP:1864708.1864753} studied an evolutional method, which can provide meaningful insights. Although many studies have been conducted in the temporal aspect of the data, most of them are retrospective, and overlook the causality of the data generation process.

\section{Conclusion}
\label{sect:conclusion}
Data arrive at real-world recommender systems rapidly and continuously. The velocity and volume at which data is coming suggest the use of streaming settings for recommendation. We delineate three challenges for recommendations under streaming settings, which are real-time updating, unknown sizes and concept shift.  We leverage these difficulties by proposing a streaming recommender system \algname in this paper. \algname manages stream inputs via a continuous-time random process.  In addition, the proposed framework is not only able to track the dynamic changes of user/item topics, but also provides real-time recommendations in a prospective way. Further, we conduct experiments on three real-world datasets and \algname significantly outperforms other state-of-the-arts.  It provides us an encouraging feedback to model data as streams.  

\scriptsize
\balance
\bibliographystyle{abbrv}
\bibliography{arxiv_2016}

\end{document}